\documentstyle[11pt]{article}

\newcommand{\be}{\begin{equation}}
\newcommand{\ee}{\end{equation}}
\newcommand{\ba}{\begin{eqnarray}}
\newcommand{\ea}{\end{eqnarray}}
\newcommand{\no}{\nonumber}

\newcommand{\dis}{\displaystyle}
\newtheorem{theorem}{Theorem}

\newtheorem{lem}{Lemma}

\begin{document}
\begin{flushright}
UT-863\\
November, 1999 \\
\end{flushright}

\bigskip

\begin{center}
{\Large \bf Generalization of Calabi-Yau/Landau-Ginzburg correspondence}
\end{center}

\bigskip

\vskip25mm

\begin{center}

Tohru Eguchi 

\bigskip

{\it Department of Physics, Faculty of Science,

\medskip

University of Tokyo, 

\medskip

Tokyo 113, Japan}

\bigskip

\bigskip

and

\bigskip

\bigskip

Masao Jinzenji

\medskip

{\it Graduate School of Mathematical Science

\medskip

University of Tokyo

\medskip

Tokyo 153, Japan}
\end{center}

\bigskip

\begin{abstract}
We discuss a possible generalization of the Calabi-Yau/Landau-Ginzburg 
correspondence to a more general class of manifolds. 
Specifically we consider the Fermat type hypersurfaces $M_N^k$: $\sum_{i=1}^N
X_i^k =0$ in 
${\bf CP}^{N-1}$ for various values of $k$ and $N$.
When $k<N$, the 1-loop beta function of the sigma model on $M_N^k$ is 
negative and we 
expect the theory to have a mass gap. However, the quantum
cohomology relation $\sigma^{N-1}=\mbox{const.}\sigma^{k-1}$ suggests that
in addition to the massive vacua there exists a remaining massless 
sector in the theory if $k>2$. We assume that this massless sector is 
described by a 
Landau-Ginzburg (LG) theory of central charge $c=3N(1-2/k)$ 
with $N$ chiral fields with $U(1)$ charge $1/k$.
We compute the topological invariants (elliptic genera) 
using LG theory and massive vacua and compare them with the geometrical data. 
We find that
the results agree if and only if $k=$even and $N$=even. 

These are the
cases when the hypersurfaces have a spin structure. Thus we find an evidence
for the geometry/LG correspondence in the case of spin manifolds.

\end{abstract}

\newpage

\section{Introduction}

It is well-known that it is possible to reproduce various geometrical data 
of Calabi-Yau manifolds by making using the orbifoldized Landau-Ginzburg 
theory.
Such a correspondence between Calabi-Yau manifold and the Landau-Ginzburg 
theory has been known for quite some time \cite{VW,Martinec} and has been
studied extensively in connection with mirror symmetry.
A simple explanation of this correspondence was provided some time ago using 
the gauged linear sigma model with a Fayet-Illiopoulos parameter $r$ 
\cite{Wittena}. In \cite{Wittena} it was shown that in the limit of large 
positive values of $r$ the non-linear sigma model on the Calabi-Yau (CY) 
manifold is recovered while 
the theory reduces to a Landau-Ginzburg (LG) type model in the limit of large
negative values of $r$. $r$ has the meaning of the size of the Calabi-Yau 
manifold.
Since the topological quantities remain invariant under a smooth variation 
of the Fayet-Illiopoulos parameter, one recovers topological invariants of
CY manifolds by studying LG theories.

In this article we would like to generalize such a 
Calabi-Yau/Landau-Ginzburg correspondence to a more general class of 
manifolds. Specifically in the following 
we consider Fermat-type hypersurfaces $M_N^k:\hskip2mm 
X_1^k+X_2^k+\cdots+X_N^k=0$ in 
${\bf CP}^{N-1}$ for various value of $k$ and $N$. $k=N$ give the 
$N-2$-dimensional Calabi-Yau manifold. When $k<N$, the hypersurface has a 
positive 1st 
Chern class $N-k$ and is a typical Fano variety. We recall that the 
(one-loop) 
beta function of the supersymmetric non-linear sigma model on a manifold $M$ 
is proportional to the minus of its 1st Chern class.  We 
may then imagine that a quantum theory on Fano variety has a non-zero 
mass gap due to its asymptotic freedom.

However, the quantum cohomology relation 
(restricted to the K\"{a}hler subring) of the manifold $M_N^k$ 
is known \cite{Collino} and given by 
\be
\sigma^{N-1}=\beta k^k\sigma^{k-1}
\label{colj}\ee
where $\beta$ is the world-sheet instanton amplitude 
($\sigma$ denotes the vacuum value of the scalar component of the vector 
multiplet in the linear $\sigma$ model. It 
corresponds to the K\"{a}hler class of $M_N^k$). (\ref{colj}) generalizes
the well-known relation for quantum cohomology ring of ${\bf CP}^{N-2}$
\be
\sigma^{N-1}=\beta.
\ee
The above equation (\ref{colj}) strongly 
suggests that 
the theory actually consists of both massive and massless sectors if $k>2$: 
(i) in the massive sector there exist $N-k$ split vacua 
corresponding to the roots of $\sigma^{N-k}=\beta k^k$ of (\ref{colj}).
(ii) in addition the theory possesses a massless sector corresponding to the 
the degenerate solution $\sigma=0$ of the relation (\ref{colj}). This 
massless sector should be described by some conformal field theory (CFT) 
in the infra-red limit. 

We have to make an identification of this
CFT describing the massless sector so that together with the massive sector 
it should reproduce the geometry of the hypersurface. A natural candidate for 
the CFT
is given by the Landau-Ginzburg theory consisting of $N$ chiral fields  
with the $U(1)$-charge $1/k$ corresponding to the defining equation
$X_1^k+X_2^k+\cdots+X_N^k=0$ of $M_N^k$. This choice seems natural since it 
gives back the standard CY-LG correspondence in the case $k=N$.

During the course of this work we became aware of an article by Witten
\cite{Wittenb} where he suggests an approach very similar to ours. 
He further discusses the
case of $k>N$, hypersurfaces of general type, where the beta function is 
positive and the direction of renormalization group flow is inverted.

In the case of Fano varieties $N>k$, field theory on the manifold 
flows in the infra-red to LG theories. If one assigns the central charge of
the free field theory on the manifold as $c_{UV}=3(N-2)$, 
then the central charge decreases in the infra-red to 
$c_{IR}=3N(1-2/k)$ of the LG-theory in accord with the theorem of 
Zamolodchikov. In the case of manifolds of 
general type $k>N$,
one flows from the LG theory in the ultra-violet with $c_{UV}=3N(1-2/k)$
to a free field theory in the
infra-red with $c_{IR}=3(N-2)$ and again the central charge decreases.

In the following by making use of the LG theory we compute 
various topological invariants for Fano and general hypersurfaces,
i.e. Euler numbers, elliptic genera for the $\sigma$ (signature) and 
$\hat{A}$-genus.
Method of computation of the elliptic genera in CFT/LG theories are 
well-known and described in  
\cite{EOTY, Vafa, Wittenc, KYY, Berglund-Henningson, Borisov}. 
We take $N$ to be even so that 
the signature and $\hat{A}$-genus of the manifold are well-defined. 
We then compare the results
with the geometrical computation of elliptic genera for the hypersurfaces. 
Geometrical method of computation may be found 
in the literature \cite{Schellekens-Warner, Wittend}.
 
It turns out that we find exact agreements between LG theory and geometry
if and only if $k$=even. Note that with $N$=even the 1st Chern class $N-k$
is divisible by 2 when $k$ is even and the manifold has a spin 
structure. Therefore we have found an evidence that the geometry/LG 
correspondence exists in a class of spin manifolds. 

We have also examined a few examples of hypersurfaces in 
a weighted projective space and again
verified the LG/geometry correspondence when the hypersurface has a spin 
structure.

\bigskip

\section{Quantum cohomology relation for hypersurfaces}

Let us start with the relation of quantum cohomology obtained by
Collino and Jinzenji \cite{Collino} for hypersurface. We first recall 
that the cohomology classes of hypersurfaces consist of two classes (1) and 
(2): elements in the class (1) are generated by the powers of the K\"{a}hler 
class $\omega$ 
and are given by $\{1,\omega,\omega^2,\cdots,\omega^{N-2}\}$. We call this 
class as the K\"{a}hler subring in the following. Elements in the class (2) 
are given by the $(p,q)$-forms at the
middle dimension, $\{{\cal O}^{p,q}, \hskip2mm p+q=N-2\}$. Let us call them as
primitive classes.
Due to the Lefschetz theorem these elements exhaust the cohomology classes 
of the hypersurface. In the quantum cohomology at the genus=0 level these two
classes are essentially decoupled from each other and one can restrict 
one's attention to the sector of K\"{a}hler subring. 

It is then possible to derive the following 
relation for the degree $k$ hypersurface ($k<N$)
\be
\omega^{N-1}=\beta k^k\omega^{k-1}.
\label{qucoh}
\ee
Here $\beta$ denotes the basic 1-instanton amplitude. 
We note that eq.(\ref{qucoh}) generalizes the well-known formula for
${\bf CP}^{N-2}$
\be
\omega^{N-1}=\beta.
\label{cp}\ee
When one uses the description of the supersymmetric linear sigma model 
\cite{Wittena}, $\omega$ is replaced by the scalar field $\sigma$ of the 
${\cal N}=2$ vector multiplet in 2-dimensions.

Quantum cohomology relation for ${\bf CP}^{N-2}$ (\ref{cp}) may be 
interpreted as the equation for the minima of a perturbed
superpotential $W=1/N\cdot\sigma^N-\beta\sigma$. In this case the quantum 
deformation corresponds to the most relevant perturbation. At each of the
$N-1$ minima 
$\{\sigma_j=\mbox{const}\cdot e^{2\pi ij/(N-1)},\hskip1mm j=0,1,2\cdots,N-2,\hskip2mm
\mbox{const}=\beta^{1/(N-1)}\}$, the second derivative of the 
superpotential does not
vanish and the theory has a mass gap.
Also in the case of $k=2$ the second derivative does not vanish at all
minima and the system has a map gap.

However, in the case of larger values of $k\ge 3$
the perturbation becomes less relevant and the degeneracy of the vacua is 
not completely resolved. The perturbed 
superpotential $W=1/N\cdot\sigma^N-
\beta k^{k-1}\sigma^k$ possesses both the $N-k$ massive vacua at 
$\{\sigma_j=\mbox{const}\cdot e^{2\pi ij/(N-k)},
\hskip1mm j=0,1,2\cdots,N-k-1,\hskip2mm
\mbox{const}=(\beta k^k)^{1/(N-k)}\}$ and the 
degenerate vacua at the origin $\sigma=0$. Thus the system has 
both massive and massless degrees of freedom.

\bigskip

\section{Landau-Ginzburg models}

We assume that the massless degrees of freedom are described by the
Landau-Ginzburg theory corresponding to the
tensor product of $N$ copies of 
${\cal N}=2$ superconformal minimal models with level $k-2$. Central
charge of this system is given by $c=3N(1-2/k)$.

Let us first check if the sum of the number of massive vacua and the 
degenerate vacua of CFT add up to the Euler number of 
the hypersurface. We may use the ${\cal N}=2$ representation theory for the 
CFT sector, however, for the computation of topological invariants a simple 
free-field realization may be used. 

Let us introduce the ratio of
theta functions
\be
Z_{LG}(\gamma)={1\over k}\sum_{a,b=0}^{k-1}n_{a,b}
\left({\theta_1\big(\displaystyle{(1-k) \over k}\gamma
+\displaystyle{b \over k}
+\displaystyle{a \tau \over k} \bigm|\tau \bigm)  \over 
\theta_1\big(\displaystyle{1 \over k}\gamma+\displaystyle{b \over k}
+\displaystyle{a \tau \over k} \bigm|\tau \bigm)}\right)^N 
\label{LG1}\ee
where $\theta_1$ is defined as usual 
\ba
&&\hskip-7mm\theta_1(z\thinspace|\thinspace\tau)=-iq^{1/8}(e^{\pi iz}-e^{-\pi iz})
\prod_{{\scriptstyle n=1}}(1-q^n)(1-q^ne^{2\pi i z})(1-q^ne^{-2\pi i z}),\\
&&\hskip-7mm q=\exp(2\pi i \tau) \no
\ea
$n_{a,b}$ is a phase factor which will be determined by imposing modular 
invariance.

We also recall the definition of $\theta_i, \hskip1mm i=2,3,4$ functions
\ba
&&\theta_2(z\thinspace|\thinspace\tau)=q^{1/8}(e^{\pi iz}+e^{-\pi iz})
\prod_{{\scriptstyle n=1}}(1-q^n)(1+q^ne^{2\pi i z})(1+q^ne^{-2\pi i z}),\\
&&\theta_3(z\thinspace|\thinspace\tau)=
\prod_{{\scriptstyle n=1}}(1-q^n)(1+q^{n-1/2}e^{2\pi i z})
(1+q^{n-1/2}e^{-2\pi i z}),\\
&&\theta_4(z\thinspace|\thinspace\tau)=
\prod_{{\scriptstyle n=1}}(1-q^n)(1-q^{n-1/2}e^{2\pi i z})
(1-q^{n-1/2}e^{-2\pi i z}).
\ea

\bigskip

{\large \bf  Euler number}\\

The above formula (\ref{LG1}) corresponds to an amplitude
\be
Z_{CFT}=\mbox{Tr}(-1)^Fe^{2\pi i\gamma J_0}q^{L_0}\bar{q}^{\bar{L}_0}
\label{ellCFT}\ee
in the superconformal field theory where $F$ is the sum 
of left 
and right-moving fermion numbers $F_L+F_R$ and $J_0$ is the zero mode of the 
$U(1)$ current in the left-handed sector. $Z_{CFT}$ becomes the Euler number 
of the target manifold when $\gamma=0$.

If we choose $\gamma=0$ in the corresponding formula of LG theory (\ref{LG1}), 
all the oscillator modes in fact 
cancel and we have contributions only from zeros modes. We set $n_{a,b}=1$. Then
we find 
\be
Z_{LG}(\gamma=0)=\displaystyle{(1-k)^N \over k}+\displaystyle{k^2-1 \over k}.
\label{Euler2}\ee
where the first term comes from the sector $a=b=0$ 
(we use l'hopital's rule in evaluating this contribution)
while each of the other sectors contributes $1/k$ to the second term.

The total number of ground states is the sum of 
eq.(\ref{Euler2}) and the number of massive vacua which is equal to $N-k$. 
Thus the LG prediction for the Euler number $\chi$ of $M_N^k$ is given by
\be
\chi=\displaystyle{(1-k)^N+k^2-1 \over k}+N-k=
\displaystyle{(1-k)^N+Nk-1 \over k}.
\label{Euler}\ee
On the other hand, the Euler number of $M_N^k$ is computed geometrically 
using the adjunction formula as
\be
c_{N-2}=\left.\displaystyle{(1+H)^N \over (1+kH)}\right|_{H^{N-2}}
\hskip-2mm \times H^{N-2}
=\displaystyle{(1-k)^N-(1-kN) \over k^2}H^{N-2}.
\ee
Here $H$ denotes the hyperplane class of $M_N^k$ 
and $c_{N-2}$ is the $(N-2)$-th Chern 
class. The symbol $|_{H^n}$ means to take the 
coefficient of the $H^n$ term. 
Integrating $c_{N-2}$ over the manifold and using $\int_{M_N^k} H^{N-2}=k$ we
recover the LG prediction (\ref{Euler}).

We may rewrite the above Euler number $\chi$ as
\be
\chi=\displaystyle{(1-k)^N+k-1 \over k}+N-1
\label{Euler3}\ee
In (\ref{Euler3}) we recognize the 2nd term as the number 
of elements in the K\"{a}hler subring and then the first term is 
identified as the number of primitive 
elements of $M_N^k$. On the other hand, in the LG description the first term 
is identified as the sum
of contributions from the untwisted sectors $a=0,b=0,\cdots,k-1$ while
the 2nd term equals the sum of contributions from the twisted sectors and 
the massive vacua. Thus we find the correspondence
\ba
&&\mbox{untwisted sector} \Longleftrightarrow \mbox{primitive classes},\\
&&\mbox{twisted sector+massive vacua} \Longleftrightarrow 
\mbox{K\"{a}hler subring}.
\ea
We recognize that the quantum deformation takes place only in the
sector of K\"{a}hler subring and the primitive classes remain intact
under quantum deformation.

\bigskip

{\large \bf Hirzebruch signature}\\

Let us now consider the Hirzebruch signature $\sigma$ and its elliptic 
generalization. First we recall that the signature is well-defined only for 
$4n$ real dimensional manifolds. Thus we take $N=$even hereafter.

In the LG formulation the elliptic genus for the signature is given by 
(\ref{LG1}) with $\gamma=1/2$,
\ba
&&\chi_{\sigma}(\tau)={1\over k}\sum_{a,b=0}^{k-1}n_{a,b}
\left({\theta_1\big(\displaystyle{(1-k) \over 2k}
+\displaystyle{b \over k}
+\displaystyle{a \tau \over k} \bigm|\tau \big)  \over 
\theta_1\big(\displaystyle{1 \over 2k}+\displaystyle{b \over k}
+\displaystyle{a \tau \over k} \Bigm|\tau \big)}\right)^N, 
\label{LG2}\\
&&={(-1)^N\over k}\sum_{a,b=0}^{k-1}n_{a,b}
\left({\theta_2\big(\displaystyle{1 \over 2k}
+\displaystyle{b \over k}
+\displaystyle{a \tau \over k} \bigm|\tau \big)  \over 
\theta_1\big(\displaystyle{1 \over 2k}+\displaystyle{b \over k}
+\displaystyle{a \tau \over k} \Bigm|\tau \big)}\right)^N.
\ea
We assume that there is no contribution to the signature from the 
massive vacua. 

First we study the modular property of (\ref{LG2}) and determine the possible
choice of phase factors $n_{a,b}$. It is known that the elliptic genus
is a modular form under the group $\Gamma_0(2)$ which leaves invariant a
fixed spin structure.
$\Gamma_0(2)$ is the subgroup of $SL(2,{\bf Z})$ of index $6$ and consists of
matrices of the form
\be
\Gamma_0(2)=\left\{\left(\begin{array}{cc} a & b \\ c & d \end{array}
\right)=\left(\begin{array}{cc} 1 & 0 \\ 0 & 1 \end{array}
\right), \mbox{ mod 2} \right\}.
\ee
It is generated by the elements $T^2,\hskip1mm ST^2S^{-1}$ where $S$ and $T$
are the standard generators of $SL(2,{\bf Z})$
\be
S=\left(\begin{array}{cc} 0 & -1 \\ 1 & 0 \end{array}\right),
\hskip3mm 
T=\left(\begin{array}{cc} 1 & 1 \\ 0 & 1 \end{array}\right).
\ee
We recall the modular properties of $\theta_1(z|\tau)$,
\ba
&&\hskip-12mm\theta_1(z\pm \tau \thinspace|\thinspace\tau)=-exp(-i\pi (\tau\pm 2z))
\theta_1(z\thinspace | \thinspace \tau),\\
&&\hskip-12mm\theta_1\Bigm({ z \over c\tau+d} \thinspace \Bigm | \thinspace 
{a\tau+b \over c\tau+d}\Bigm)=(-i)^{1/2}(c\tau+d)^{1/2}\exp\Big(\displaystyle
{i\pi c z^2 \over c\tau+d}\Big)\thinspace
\theta_1(z\thinspace |\thinspace \tau).
\ea

It is easy to see that under $T^2:\tau \rightarrow \tau +2$, 
(\ref{LG2}) transforms as
\be
\chi_{\sigma}(\tau+2)={1 \over k}
\sum_{a,b=0}^{k-1}n_{a,b-2a}
\left({\theta_1\big(\displaystyle{(1-k) \over 2k}
+\displaystyle{b \over k}
+\displaystyle{a \tau \over k} \bigm|\tau \big)  \over 
\theta_1\big(\displaystyle{1 \over 2k}+\displaystyle{b \over k}
+\displaystyle{a \tau \over k} \Bigm|\tau \big)}\right)^N.
\ee
 On the other hand, under $ST^2S^{-1}:
\tau \rightarrow \tau/(-2\tau+1)$, $\chi_{\sigma}(\tau)$
transforms as
\ba
\hskip-20mm \chi_{\sigma}(\dis{\tau\over -2\tau+1})
&\hskip-3mm=&\hskip-3mm{1\over k}\sum_{a,b=0}^{k-1}n_{a,b}
\left({\theta_1\big(\dis{(1-k) \over 2k}
+\dis{b \over k}
+\dis{a\over k}\dis{\tau \over -2\tau+1}
\bigm|\dis{\tau \over -2\tau+1}\bigm)  \over 
\theta_1\big(\displaystyle{1 \over 2k}+\displaystyle{b \over k}
+\displaystyle{a\over k}\displaystyle{\tau \over -2\tau+1}
\bigm|\displaystyle{\tau \over -2\tau+1}\bigm)}\right)^N \\
&=& {1\over k}\sum_{a,b=0}^{k-1}
n_{a,b}\exp\left[i\pi N({2a' \over k}+{b \over k}
+{2-k\over 2k}+\tau)\right]\\
&&\hskip10mm \times\left({\theta_1\big(\displaystyle{(1-k) \over 2k}
+\displaystyle{b \over k}
+\displaystyle{a'\over k}\tau+\tau \bigm|\tau\big)  \over 
\theta_1\big(\displaystyle{1 \over 2k}+\displaystyle{b \over k}
+\displaystyle{a'\over k}\tau 
\bigm|\tau \big)}\right)^N\hskip-2mm, \hskip1mm a'=a-2b-1 \\
&=&  {i^N \over k}\sum_{a,b=0}^{k-1}n_{a+2b+1,b}
\left({\theta_1\big(\displaystyle{(1-k) \over 2k}
+\displaystyle{b \over k}
+\displaystyle{a \tau \over k} \bigm|\tau \bigm)  \over 
\theta_1\big(\displaystyle{1 \over 2k}+\displaystyle{b \over k}
+\displaystyle{a \tau \over k} \bigm|\tau \bigm)}\right)^N\hskip-2mm . 
\ea
We find that there are two natural choices for the phase factor $n_{a,b}$
\ba
&&\begin{array}{ll}(i) &  n_{a,b}=1,\\
(ii)& n_{a,b}=(-1)^a. \end{array}
\ea
Correspondingly $\chi_{\sigma}$ transforms under $T^2$ and $ST^2S^{-1}$ as
\ba
&&\begin{array}{lllrl}(i) & \chi_{\sigma} & \longrightarrow & 
\chi_{\sigma}, & (-1)^{N/2}\chi_{\sigma}\\
(ii) & \chi_{\sigma} & \longrightarrow &\chi_{\sigma},
& -(-1)^{N/2}\chi_{\sigma}.
\end{array}
\label{modular2}\ea
If we examine the modular property of known formula for the elliptic genus,
for instance, of $K_3$ surface \cite{EOTY}, 
we find that the choice $(ii)$ gives the 
correct transformation law. Thus we adopt $n_{a,b}=(-1)^a$ hereafter in this 
section,
\be
\chi_{\sigma}(\tau)=-{(-1)^{N \over 2}\over k}\sum_{a,b=0}^{k-1}(-1)^a
\left({\theta_2\big(\displaystyle{1 \over 2k}
+\displaystyle{b \over k}
+\displaystyle{a \tau \over k} \bigm|\tau \bigm)  \over 
\theta_1\big(\displaystyle{1 \over 2k}+\displaystyle{b \over k}
+\displaystyle{a \tau \over k} \bigm|\tau \bigm)}\right)^N.
\label{LG3}
\ee
Here we also have 
adjusted the overall sign so that (\ref{LG3}) agrees with the 
standard convention for the signature.

Let us first test (\ref{LG3}) by comparing its ground state multiplicity
with the classical value of the signature. In the case of $N=4$ 
(complex surface), for instance, 
it is easy to compute geometrically the 
signature of $M_4^k$ 
if we use (\ref{ellhyper}),(\ref{powersum}) of the next section,
\be
\sigma(M_4^k)={1\over 3}(4-k^2)k.
\ee
$\sigma(M_4^4)=-16$ is the well-known value for the $K_3$ surface.

On the other hand, the degeneracy of the lowest state
in LG theory (\ref{LG3}) is given by
\ba
&&\hskip-6mm \sigma_{LG}(M_N^k)
=-{1 \over k}\left(\sum_{b=0}^{k-1}
\left(\displaystyle{1+e^{-2\pi i (1/2k+b/k)}\over 1-e^{-2\pi i (1/2k+b/k)}}
\right)^N+k\sum_{a=1}^{k-1} (-1)^a\right)
\ea
where the first sum comes from the untwisted sector and the second from
the twisted sectors. Note that the contributions from twisted 
sectors cancel when $k=$odd while they add up to $+1$ when $k=$even. 
We may rewrite the first sum as
\be
-\sum_{n=1}^{N-1}(-1)^n \times
\Bigm(t^{{1 \over k}}\Big(1+t^{{1 \over k}}+t^{{2 \over k}}+
\cdots+t^{{k-2 \over k}}\Big)\Bigm)^N\hskip-1mm
\left|_{\,\displaystyle{t^{n}}}\right..
\label{sigLG}\ee
(\ref{sigLG}) is an alternating sum of coefficients of integer powers of $t$.
In the case of $N=4$, for instance, after a simple calculation we find that
(\ref{sigLG}) is equal to
\be
{1 \over 3}(4-k^2)k-1.
\ee 
Thus
\ba
\sigma_{LG}=\sigma, \hskip5mm k=\mbox{even}
\ea
while $\sigma_{LG}=\sigma-1, \hskip1mm k=\mbox{odd}$.
Therefore the prediction of the LG theory agrees 
with the geometry only when $k=$even. When $k=$odd, there is a 
missing factor of 1 in $\sigma_{LG}$. 

More precisely we may identify the above coefficients in (\ref{sigLG}) as the 
the Hodge numbers. In the case of $N=4$
\be
c_n=\Bigm(t^{{1 \over k}}\Big(1+t^{{1 \over k}}+t^{{2 \over k}}+
\cdots+t^{{k-2 \over k}}\Big)\Bigm)^4\hskip-1mm
\left|_{\,\displaystyle{t^{n}}}\right.=h^{3-n,n-1}-\delta_{n,2}, 
\hskip2mm n=1,2,3.
\ee
Here a factor 1 is subtracted from $h^{1,1}$ in order to eliminate
the contribution from the K\"{a}hler class.
In the case $k=$even this factor 1 together with
an additional factor 1 from the twisted sectors adds up to 2 and 
one obtains
\be
\sigma_{LG}=2+2h^{2,0}-h^{1,1}
\label{sigind}\ee
in agreement with the index theorem. In the case $k=$odd, however,
twisted sectors cancel and LG theory fails to reproduce (\ref{sigind}). 
One can check that this phenomenon occurs for all values of $N$.
 
We may further test the agreement of the LG theory with the geometry
at excited levels, in particular in the case of $k=$even.
Unfortunately, it is not easy to evaluate the expression (\ref{LG3})
exactly except for the trivial case $k=1$ where $\chi_{\sigma}(\tau)$ 
vanishes 
identically,
\be
k=1: \hskip4mm \chi_{\sigma}(\tau)=0, \hskip4mm \mbox{any }N.
\label{LGk1}\ee

In the case $k=$even, however, we may repeatedly use the addition formula 
of the $\wp$ function and evaluate (\ref{LG3}) in a closed form. 
Details are described in the appendix. Computation in the case $k=2$ is 
straightforward and is given by
\ba
&&k=2: \hskip4mm \chi_{\sigma}(\tau)= \left\{\begin{array}{cc}
0 & \mbox{ $\displaystyle{N\over 2}$  is even} \\
 & \\
2 &  \mbox{ $\displaystyle{N \over 2}$ is odd}\end{array}\right.
\label{LGk2}\ea
This is in accord with the fact that when $k=2$, central charge
of the CFT vanishes ($c=N(1-2/k)=0$) and the theory does not possess 
excited states.

After somewhat lengthy computations we obtain expressions for 
$\chi_{\sigma}(\tau)$ for $k=4$ and $k=6$ with various values of $N$
\ba
\hskip-10mm &&k=4: 
\hskip3mm \chi_{\sigma}(\tau)=\left\{\begin{array}{lc}
-8\left(\displaystyle{(\theta_3(0|\tau))^2 \over (\theta_4(0|\tau))^2}+
\displaystyle{(\theta_4(0|\tau))^2 \over (\theta_3(0|\tau))^2}\right)  
& \hskip-25mm
N=4 \\
\null & \\
100,  & \hskip-25mm N=6 \\
\\
-288\left(\displaystyle{(\theta_3(0|\tau))^2 \over (\theta_4(0|\tau))^2}+
\displaystyle{(\theta_4(0|\tau))^2 \over (\theta_3(0|\tau))^2}\right),  
& \hskip-25mm
N=8 \\
\\
4\left(681+80\displaystyle{(\theta_3(0|\tau))^4 \over (\theta_4(0|\tau))^4}
+80\displaystyle{(\theta_4(0|\tau))^4 \over (\theta_3(0|\tau))^4}\right), 
& \hskip-25mm
N=10\\
\\
-8\left(\displaystyle{(\theta_3(0|\tau))^2 \over (\theta_4(0|\tau))^2}+
\displaystyle{(\theta_4(0|\tau))^2 \over (\theta_3(0|\tau))^2}\right)
\left(1193+16\displaystyle{(\theta_3(0|\tau))^4 \over (\theta_4(0|\tau))^4}
+16\displaystyle{(\theta_4(0|\tau))^4 \over (\theta_3(0|\tau))^4}\right), 
& N=12\\
\end{array}\right.
\label{LGk4}\ea
\ba
\hskip-12mm &&k=6:  
\hskip3mm \chi_{\sigma}(\tau)=\left\{\begin{array}{lc}
-32\left(\displaystyle{(\theta_3(0|\tau))^2 \over (\theta_4(0|\tau))^2}+
\displaystyle{(\theta_4(0|\tau))^2 \over (\theta_3(0|\tau))^2}\right), 
& N=4 \\
\\
2\left(419+16\displaystyle{(\theta_3(0|\tau))^4 \over (\theta_4(0|\tau))^4}+
16\displaystyle{(\theta_4(0|\tau))^4 \over (\theta_3(0|\tau))^4}\right),  
& N=6 \\
\\
-6272\left(\displaystyle{(\theta_3(0|\tau))^2 \over (\theta_4(0|\tau))^2}+
\displaystyle{(\theta_4(0|\tau))^2 \over (\theta_3(0|\tau))^2}\right),  & N=8 
\end{array}\right..
\label{LGk6}\ea
We find that the formulas for $k=2,4,6$ all agree with the geometrical data
presented in the next section.

On the other hand in the case of odd $k$ we can not evaluate 
(\ref{LG3}) in a closed form but only obtain its $q$-expansion. We list the
formulas for $k=3$ and $5$
\ba
&&\hskip-10mm k=3:  \hskip3mm \chi_{\sigma}(\tau)=\left\{\begin{array}{lc}
-6-48q-240q^2-912q^3+\cdots,  & N=4 \\
\\
18+216q+1512q^2+7848q^3+\cdots,  & N=6 \\
\\
-54-864q-7776q^2-50976q^3+ \cdots,  & N=8 
\end{array}\right.
\label{LGk3}\ea
\ba
\hskip-12mm&&k=5:  \hskip1mm \chi_{\sigma}(\tau)=\left\{\begin{array}{lc}
-36-464q-3472q^2-19392q^3+\cdots, & N=4  \\
\\
340+6600q+70680q^2+548640q^3+\cdots,  & N=6 \\
\\
-3220-83360q-1162400q^2+\cdots,  & N=8.
\end{array}\right.
\label{LGk5}\ea

By comparing these results with the geometrical data 
we conclude that the LG predictions reproduce the elliptic genus 
for signature
exactly if and only if $k=$ even. 
We note 
that the hypersurface $M_N^k$ with $N=$even is a spin manifold 
when $k=$even and hence we find the agreement of LG theory with geometry 
when the target space is a spin manifold. 

Condition of spin manifold seems 
natural since the signature is closely related to the $\hat{A}$ genus 
which exists only for the spin manifold.
In the case of Calabi-Yau manifolds the elliptic genera for 
$\sigma$ and $\hat{A}$ 
are in fact given by the same modular function evaluated at different 
values of its argument. 

We note that in the previous section Euler number of the manifold 
was correctly computed using LG theory without any condition for $N$ and $k$.
In section 5 we find that the elliptic genus for $\hat{A}$ (which exists
only for $N=$even and $k=$even) can also be correctly reproduced by LG 
theory. Thus LG theory works at
\be
\left\{\begin{array}{ll}
\chi\hskip1mm (\mbox{exists for any $N$ and $k$}) & \mbox{LG works for any $N$, \hskip1mm
$k$} \\
\sigma\hskip1mm(\mbox{exists for $N=$even}) & \mbox{LG works for $N=$even and 
$k=$even}\\
\hat{A}\hskip1mm (\mbox{exists for $N=$even and $k=$even})
&\mbox{LG works for $N=$even and $k=$even}.
\end{array}
\right.
\ee

\section{geometrical computation}

Let us now turn to the geometrical evaluation of the elliptic genus 
in order to
provide data to test LG predictions. 
Outline of calculation goes as follows. 

Elliptic genus for the loop-space signature operator on 
a manifold $M$ of complex dimension $d$ is defined as \cite{Och,LS}
\be
\sigma({\cal L}M)=2^d\int_Mch({\bf E}_{q})L(M)
\ee
where $ch({\bf E}_q)$ is the Chern character for a bundle ${\bf E}_{q}$
\be
{\bf E}_q=(\otimes_{n\ge 1}(\Lambda_{q^n}TM))\otimes
(\otimes_{n\ge 1}(\Lambda_{q^n}T^*M))\otimes(\otimes_{n\ge 1}S_{q^n}
(TM\oplus T^*M)).
\ee
Here $\Lambda_{q^n}(TM)$ means, for instance, $1+q^nTM+q^{2n}\Lambda (TM)^2+
\cdots$ and 
$S(\Lambda)$ stands for the 
(anti-)symmetrization of tensor product of
tangent bundles $TM$.
$T^*M$ is the dual of $TM$. 

$L(M)$ is the Hirzebruch L-polynomial
\be
L(M)=\prod_{{\scriptstyle i=1}}^{{\scriptstyle d}}
{(\displaystyle{u_i \over 2}) \over \tanh(\displaystyle{u_i \over 2})}
\ee
where the tangent bundle $TM$ is split into a sum
of line bundles with 1st Chern classes $\{u_i,i=1,\cdots,d\}$.
Hirzebruch polynomial may be expanded as
\be
{\displaystyle{u\over 2}\over \tanh\displaystyle{u \over 2}}
=\exp\Big(\sum_{j=1}^{\infty}(-1)^{j-1}\displaystyle{(2^{2j-1}-1)\over j\cdot (2j)!}
B_ju^{2j}\Big)
\ee
where $B_{j}$'s are the Bernoulli numbers.

Next by introducing an auxiliary function
\be
f(q,z)=\prod_{{\scriptstyle n=1}}\displaystyle{(1+e^{2\pi iz}q^{n})
(1+e^{-2\pi iz}q^{n})\over (1-e^{2\pi iz}q^n)(1-e^{-2\pi iz}q^n)}
\ee
we define
\be
F_{2j}(q)={1 \over 2}({1 \over 2\pi i}{d \over dz})^{2j}\log f(q,z)
\left|_{z=0}\right.
\hskip2mm j=1,2,3\cdots 
\ee
We then have an identity
\be
\exp({u \over 2\pi i}{d \over dz})
f(q,z)|_{z=0}=f(q)\exp\left(\sum_{j=1}^{\infty}
{2 \over (2j)!}F_{2j}(q)u^{2j}\right)
\ee
where
\be
f(q)=f(q,z)|_{z=0}.
\ee

Then the elliptic genus is expressed as
\be
\sigma({\cal L}M)=\int_M \left(2f(q)\right)^{d}
\exp\left(\sum_{j=1}^{\infty}
{2 \over (2j)!}\left((-1)^{j-1}(2^{2j-1}-1){B_{j}\over 2j}+F_{2j}(q)\right)
\Big(\sum_{i=1}^{{d}}u_i^{2j}\Big)\right).
\label{ellhyper}\ee

In the case of the degree-$k$ hypersurface in ${\bf CP}^{N-1}$ 
Potryagin classes are given by
\be
\sum_{n=0}^{\infty}(-1)^np_n=\prod_{{\scriptstyle i=1}}^{\scriptstyle {N-2}}
(1-u_i^2)
={(1-H^2)^N\over (1-k^2H^2)}
\label{pont}\ee
where $H$ is the hyperplane class. 
Thus by taking the logarithm of (\ref{pont}) we find 
\be
\sum_{i=1}^{N-2}u_i^{2j}=(N-k^{2j})H^{2j}.
\label{powersum}\ee

It is then easy to generate $q$-series for $\sigma({\cal L}M)$ for various
values of $k$ and $N$. It turns out that again in the case $k=$even it is 
possible to obtain a closed formula for the elliptic genus. We have evaluated
them in the case of $k=2,4,6$ and
reproduced exactly the results of the LG theory 
(\ref{LGk2}),(\ref{LGk4}),(\ref{LGk6}). Details are given in the appendix.

On the other hand in the case $k=$odd we obtain only the $q$-series.
We present the results of $k=1,3,5$, 
\ba
k=1: \hskip3mm \sigma({\cal L}M)=\left\{\begin{array}{lc}
1+32q+256q^2+1408q^3+\cdots, & N=4 \\
\\
1+96q+2304q^2+28800q^3+\cdots, & N=6 \\
\\
1+192q+9216q^2+213248q^3 +\cdots, & N=8\\
\end{array}\right.
\label{sigk1}\ea

\ba
k=3: \hskip3mm \sigma({\cal L}M)=\left\{\begin{array}{lc} 
-5-160q-1280q^2-7040q^3+\cdots., & N=4 \\
\\
19-224q-5376q^2-67200q^3+\cdots,  & N=6\\
\\
-53-1984q-29696q^2-455936q^3+\cdots, & N=8\\
\end{array}\right. 
\label{sigk3}\ea

\ba
k=5: \hskip3mm \sigma({\cal L}M)=\left\{\begin{array}{lc}
-35-1120q-8960q^2-49280q^3+ \cdots,  & N=4 \\
\\
341+2016q+48384q^2+604800q^3\cdots,  & N=6 \\
\\
-3219-101952q-764928q^2-3134208q^3+\cdots, & N=8 \\
\end{array}\right.
\label{sigk5}\ea
(\ref{sigk1}),(\ref{sigk3}),(\ref{sigk5}) differ completely from those of
the LG theory (\ref{LGk1}),(\ref{LGk3}),(\ref{LGk5}).

\section{$\hat{A}$ genus}

Now let us turn to the discussion of the $\hat{A}$ genus. $\hat{A}$ is defined
for real $4n$ dimensional spin manifolds and hence we consider hypersurfaces 
with even $N$ and $k$. In the LG formulation the elliptic genus for
$\hat{A}$ is defined by
\ba
&&\chi_{\hat{A}}(\tau)={-i(-1)^{N \over 2}\over k}\sum_{a,b=0}^{k-1}(-1)^{a+b}
\left({\theta_3\big(\displaystyle{\tau+1 \over 2k}
+\displaystyle{b \over k}
+\displaystyle{a \tau \over k} \bigm|\tau \bigm)  \over 
\theta_1\big(\displaystyle{\tau+1 \over 2k}+\displaystyle{b \over k}
+\displaystyle{a \tau \over k} \bigm|\tau \bigm)}\right)^N.
\label{LG4}
\ea
Basically $\theta_2$ in $\chi_{\sigma}$ is replaced by $\theta_3$ in
$\chi_{\hat{A}}$. The sign $(-1)^{a+b}$ is fixed again by the consideration
of modular invariance.

On the other hand the geometrical definition of the $\hat{A}$ elliptic genus 
is given by
\be
\hat{A}({\cal L}M)=q^{-{d \over 8}}\int_M ch(\tilde{{\bf E}}_q)\hat{A}(M)
\ee
where $\tilde{{\bf E}}_q$ is the bundle
\be
\tilde{{\bf E}}_q=(\otimes_{n\ge 1}(\Lambda_{q^{n-{1\over 2}}}TM))\otimes
(\otimes_{n\ge 1}(\Lambda_{q^{n-{1\over 2}}}T^*M))
\otimes(\otimes_{n\ge 1}S_{q^n}
(TM\oplus T^*M)).
\ee
We now have a half-integer moding for fermionic contributions corresponding to
the NS boundary condition. $\hat{A}(M)$ is the classical $\hat{A}$ genus
given by
\be
\hat{A}(M)=\prod_{{\scriptstyle i=1}}^{{\scriptstyle {{d}}}}
{\displaystyle{u_i\over 2} \over \sinh\displaystyle{u_i\over 2}}.
\ee
where $d$ is the complex dimension of the manifold.
$\hat{A}$ polynomial is expanded as
\be
{\displaystyle{u\over 2}\over \sinh\displaystyle{u \over 2}}
=\exp\Big(\sum_{j=1}^{\infty}(-1)^{j-1}\displaystyle{1\over 2j\cdot (2j)!}
B_ju^{2j}\Big).
\ee

We next introduce an auxiliary function
\be
h(q,z)=\prod_{{\scriptstyle n=1}}\displaystyle{(1+e^{2\pi iz}q^{n-{1\over 2}})
(1+e^{-2\pi iz}q^{n-{1\over 2}})\over (1-e^{2\pi iz}q^n)(1-e^{-2\pi iz}q^n)}
\ee
and define
\be
H_{2j}(q)={1 \over 2}({1 \over 2\pi i}{d \over dz})^{2j}\log h(q,z)
\left|_{z=0}\right.
\hskip2mm j=1,2,3\cdots .
\ee
We then have an identity
\be
\exp({u \over 2\pi i}{d \over dz})
h(q,z)|_{z=0}=h(q)\exp\left(\sum_{j=1}^{\infty}
{2 \over (2j)!}H_{2j}(q)u^{2j}\right)
\ee
where
\be
h(q)=h(q,z)|_{z=0}.
\ee
Then the elliptic genus is represented as
\be
\hat{A}({\cal L}M)=q^{-{d\over 8}}\int_M h(q)^{d}
\exp\left(\sum_{j=1}^{\infty}
{2 \over (2j)!}\left((-1)^{j-1}{B_{j}\over 4j}+H_{2j}(q)\right)
\Big(\sum_{i=1}^{{d}}u_i^{2j}\Big)\right).
\label{ahathyper}\ee

We have evaluated $\hat{A}$ genus for LG theory and 
compared the results from geometry. We have checked the exact agreements
between them for $k=2,4,6$ and various even values of $N$.

\section{discussions}

In this article we have discussed a possible generalization of
Calabi-Yau/Landau-Ginzburg correspondence to a more general class of manifolds.
We have considered the case of degree-$k$ hypersurfaces $M_N^k$ in the 
complex projective space ${\bf CP}^{N-1}$. 
$M_N^k$ has a positive, zero and negative 1st Chern class depending on 
$N>k$,$N=k$ and $N<k$, respectively. 
In all these cases we have found that the CFT/LG system always predicts the
correct topological invariants when $k=$even and $M_N^k$'s are spin manifolds. 
This fulfills
our expectation that smooth quantum deformation or renormalization flow 
preserves the topological characteristics of the manifolds. On the other hand,
when $k=$odd and manifold does not have spin structure, LG predictions for
the signature are 
in disagreement with the geometry.
Thus we have found some evidence that the CFT/geometry correspondence may  
exist in the case of spin manifolds.

In order to test this conjecture we have considered the following example 
of hypersurfaces in weighted projective space 
\be
\tilde{M}^k: X_1^{2k}+X_2^{2k}+X_3^k+X_4^k=0.
\ee
These surfaces are discussed in the literature \cite{MP,Nag}.
First Chern class of $\tilde{M}^k$ is given by $3-k$ and $\tilde{M}^3$ 
is a $K_3$ surface. We have studied topological invariants of
$\tilde{M}^k$ in the case of $k=1,3,5$ for which the
hypersurfaces have a spin structure and $k=2,4$ which are not spin manifolds. 

We have computed the elliptic genus for the signature and have checked that
in the case of $k=1,3,5$ the LG prediction for the lower order q-expansion 
coefficients coincide exactly those of geometry. On the other hand, 
LG theory disagree with geometry in the case $k=2,4$.
This example gives some additional evidence for our conjecture.

After finishing our computations we have found a brief remark 
in \cite{Wittena} (section 3) on the relevance of spin manifolds:
there it is pointed out that in the case of non-vanishing first Chern class 
$U(1)_R$-symmetry
of the theory becomes anomalous and the amplitude (\ref{ellCFT}) 
is no longer a 
topological invariant. However, there is a remaining discrete 
$U(1)_R$-symmetry 
$Z_{c_1}$ and (\ref{ellCFT}) is still well-defined for 
$\gamma=\mbox{integer}/c_1$. 
Thus if
$c_1$ is even and the manifold has a spin structure, the amplitude is
well-defined at $\gamma=1/2$ which corresponds to the Hirzebruch 
signature.

Our results are quite consistent with this remark: 
the LG computation works for any value of $k$ in the case of Euler number 
($\gamma=0$), however, it works only for $k=$even in the case of signature.

It will be interesting to study the geometry/LG correspondence in more 
general examples of manifolds than hypersurfaces.

\bigskip

We would like to thank E.Witten and C.Vafa for their communication
on the possible absence of mass gap in the theory on manifolds with 
non-analytic classes in connection with the Virasoro conjecture of
quantum cohomology in 1997.
We also thank J. Hashiba and M. Naka for discussions.
Research of T.E. is partly supported by the research fund for the Special 
Priority Area N0.707, Japan Ministry of Education.
Research of M.J. is supported by JSPS post-doctoral fellowship.

\bigskip

\section*{Appendix A: Evaluation of the elliptic genus}

In this appendix, we assume that both $N$ and $k$ are even 
integers.
We first recall the definition of the auxiliary function $f(q,z)$
\begin{eqnarray}
&&f(q,z)=\prod_{j=1}^{\infty}\frac{(1+e^{2\pi iz}q^{j})
(1+e^{-2\pi iz}q^{j})}{(1-e^{2\pi iz}q^{j})
(1-e^{-2\pi iz}q^{j})}=\tan(\pi z)\frac{\theta_{2}(z|\tau)}
{\theta_{1}(z|\tau)},\no\\
&&f(q,0)=\pi \frac{\theta_{2}(0|\tau)}
{\theta_{1}'(0|\tau)}.
\label{theta0}
\end{eqnarray}
By using 
\begin{eqnarray}
{\theta_{1}'(0|\tau)}=
\pi\theta_{2}(0|\tau)\theta_{3}(0|\tau)\theta_{4}(0|\tau).
\label{theta1}
\end{eqnarray}
we find $f(q,0)=1/
(\theta_{3}(0|\tau)\theta_{4}(0|\tau))$.
We start from the geometrical formula for
the elliptic genus given in section 4,
\begin{eqnarray}
&&\sigma({\cal L}M_{N}^{k}) \no \\
&&=(2f(q,0))^{N-2}
\int_{M_{N}^{k}}
\exp\left(\sum_{j=1}^{\infty}\frac{2}{(2j)!}
\left((-1)^{j-1}(2^{2j-1}-1)
\frac{B_{j}}{2j}+F_{2j}(q)\right)(N-k^{2j})H^{2j}\right),\no\\
&& \\
&& F_{2j}(q)={1\over 2}(\frac{1}{2\pi i})^{2j}
\frac{d^{2j}}{dz^{2j}}\log(f(q,z))|_{z=0}.
\label{taylor}
\end{eqnarray}
Since $F_{2j+1}(q)=0\;\;(j\geq 0)$, we may sum the series
in the exponent 
\ba
&&\sum_{j=1}^{\infty}{1 \over (2j)!}F_{2j}(q)H^{2j}=
{1\over 2}\left(\log f(q,{H \over 2\pi i})-\log f(q,0)\right),\\
&&\sum_{j=1}^{\infty}{1 \over (2j)!}F_{2j}(q)k^{2j}H^{2j}=
{1\over 2}\left(\log f(q,{kH\over 2\pi i})-\log f(q,0)\right).
\ea
Sum over the Bernoulli numbers reproduces the Hirzebruch polynomial
\ba
&&\sum_{j=1}^{\infty}\frac{1}{(2j)!}
(-1)^{j-1}(2^{2j-1}-1)
\frac{B_{j}}{j}H^{2j}=\log\left({{H \over 2} 
\over \tanh{H \over 2}}\right), \\
&&\sum_{j=1}^{\infty}\frac{1}{(2j)!}
(-1)^{j-1}(2^{2j-1}-1)
\frac{B_{j}}{j}k^{2j}H^{2j}=\log\left({{kH \over 2} 
\over \tanh{kH \over 2}}\right).
\ea
We can then rewrite $\sigma({\cal L}M_N^k)$ as
\ba
&&\sigma({\cal L}M_N^k)=(2f(q,0))^{N-2}\int_{M_N^k} \Big({{H \over 2} 
\over \tanh{H \over 2}}\Big)^N
\Big({\tanh{kH \over 2} \over {kH \over 2}} \Big)
\Big({f(q,{H\over 2\pi i})\over
f(q,0)}\Big)^N{f(q,0) \over f(q,{kH\over 2\pi i})}\no \\
&&={(-i)^{N-1} \over 2kf(q,0)}\int_{M_N^k} H^{N-1}
\Big(\frac{\theta_{2}\big({H \over 2\pi i}|\tau\big)}{\theta_{1}
\big({H \over 2\pi i}|\tau\big)}\Big)^{N}
\frac{\theta_{1}\big({kH \over 2\pi i}|\tau\big)}
{\theta_{2}\big({kH \over 2\pi i}|\tau\big)}.
\ea
Since the integral $\int_{M_N^k}$ picks up the coefficient of the 
$H^{N-2}$ term 
in the product of $\theta$ functions and $\int_{M_N^k} H^{N-2}=k$,
we can express $\sigma({\cal L}M_N^k)$ as
\begin{lem}
\begin{eqnarray}
\sigma({\cal L}M_{N}^{k})\hskip-3mm&=&\hskip-3mm \frac{(-i)^{N-1}}{2}
\theta_{3}(0|\tau)\theta_{4}(0|\tau)\oint_{C_{z=0}}\frac{dz}{2\pi i}
\Big(\frac{\theta_{2}({z \over 2\pi i}|\tau)}{\theta_{1}
({z \over 2\pi i}|\tau)}\Big)^{N}
\frac{\theta_{1}({kz \over 2\pi i}|\tau)}
{\theta_{2}({kz \over 2\pi i}|\tau)},
\no\\
&=&\hskip-3mm\frac{(-i)^{N-1}}{2}
\theta_{3}(0|\tau)\theta_{4}(0|\tau)\oint_{C_{z=0}}dz
\Big(\frac{\theta_{2}(z|\tau)}{\theta_{1}
(z|\tau)}\Big)^{N}
\frac{\theta_{1}(kz|\tau)}{\theta_{2}(kz|\tau)}.
\label{eva}
\end{eqnarray}
\end{lem}
Here the integration contour $\oint_{C_{z=0}}$ circles around the origin.
If we use the relation between the theta 
functions and the Weierstrass ${\cal P}$ function, 
\begin{equation}
\Big(\frac{\theta_{2}(z|\tau)}{\theta_{1}(z|\tau)}\Big)^{2}
=\Big(\frac{\theta_{2}(0|\tau)}{\theta_{1}'(0|\tau)}\Big)^{2}
({\cal P}(z|\tau)-e_{1}
(\tau)),
\end{equation}
we can rewrite
(\ref{eva}) as
\begin{eqnarray}
\sigma({\cal L}M_{N}^{k})=\frac{(-i)^{N-1}}{2\pi^{N-1}}
\frac{1}{(\theta_{3}(0|\tau)\theta_{4}(0|\tau))^{N-2}}
\oint_{C_{z=0}}dz\frac{({\cal P}(z|\tau)-e_{1}
(\tau))^{\frac{N}{2}}}{({\cal P}(kz|\tau)-e_{1}
(\tau))^{\frac{1}{2}}}.
\label{lem1}
\end{eqnarray}
$e_1$ denotes one of the zeros of the cubic equation representing the
elliptic curve $y^2=4x^3-g_2x-g_3=4(x-e_1)(x-e_2)(x-e_3)$.
Then the task of evaluating $\sigma({\cal L}M_N^k)$ is reduced
to finding zeros of the
function ${\cal P}(kz|\tau)-e_1(\tau)$ in the $z$-plane. 

Let us introduce the function $X(z)$
\begin{eqnarray}
&&X(z)\equiv {\cal P}(z)-e_{1},\label{Xdef}\\
&&\frac{d}{dz}X(z)=\frac{d}{dz}{\cal P}(z).
\end{eqnarray}
($\tau$ dependence is suppressed). We recall the addition theorem of the ${\cal P}$ function
\be
{\cal P}(2z)={\Big(6{\cal P}(z)^2-{g_2^2\over 2}\Big)^2 \over 4
\Big(4{\cal P}(z)^3-
g_2{\cal P}(z)-g_3\Big)}-2{\cal P}(z).
\ee
The coefficient functions $g_2$,$g_3$ of the elliptic curve 
are related to 
$\theta$--constants as
\be
g_2={2 \over 3}\pi^4(u^8+v^8+w^8), \hskip4mm
g_3={4 \over 27}\pi^6(v^4-w^4)(2u^8+v^4w^4)
\ee
where $u,v,w$ are defined by
\be
u=\theta_3(0|\tau), \hskip2mm v=\theta_4(0|\tau),\hskip2mm
w=\theta_2(0|\tau).
\ee
We can evaluate the elliptic genus in the case $k=2$ by using the above
addition formula. If we want to compute the genus for $k=4$, we have to
use the addition formula once more and 
express ${\cal P}(4z)$ in terms of ${\cal P}(z)$. If we represent the
resulting expression in terms of $X(z)$, we have
\begin{lem}
\begin{equation}
X(4z)=16\pi^{4}\frac{(uv)^{4}}
{(\frac{d}{dz}X(z))^{2}}
\cdot\frac{\Big(R_{4}(X(z))\cdot S_{4}(X(z))\Big)^{2}}
{\Big(R_{4}'(X(z))\cdot
S_{4}(X(z))-R_{4}(X(z))\cdot S_{4}'(X(z))\Big)^{2}}.
\label{lem2}
\end{equation}
\end{lem}
Here $R,S$ are degree-4 polynomials of $X$
\begin{eqnarray}
&&R_{4}(X)=X^{4}-4\pi^{2}u^{2}
v^{2}X^{3}-
\pi^{4}(4u^{6}v^{2}+
4u^{2}v^{6}+
2u^{4}v^{4})X^{2}-4\pi^{6}u^{6}v^{6}X+
\pi^{8}u^{8}v^{8}, \no \\
&& \\
&&S_{4}(X)=
X^{4}+4\pi^{2}u^{2}
v^{2}X^{3}+
\pi^{4}(4u^{6}v^{2}+
4u^{2}v^{6}-
2u^{4}v^{4})X^{2}+4\pi^{6}u^{6}v^{6}X+
\pi^{8}u^{8}v^{8},\no \\
&&
\end{eqnarray}
and $'$ means the derivative in $X$.
We then have
\begin{eqnarray}
&&\hskip-4mm \sigma({\cal L}M_{N}^{4})=
\frac{(-i)^{N-1}}{8
\pi^{N+1}(uv)^{N}}
\oint_{C}dX\cdot X^{\frac{N}{2}}\cdot{R'_4(X)S_4(X)-R_4(X)S'_4(X)\over
R_4(X)S_4(X)}   \\
&&=\frac{(-i)^{N-1}}{8
\pi^{N+1}(uv)^{N}}
\oint_{C}dX\cdot X^{\frac{N}{2}}\cdot\sum_{i=1}^{4}
\big(\frac{1}{X-\alpha_{i}}-\frac{1}{X-\beta_{i}}\big)\\
&&=-\frac{(-i)^{N}}{4\pi^{N}(uv)^{N}}
\left(\sum_{i=1}^4 \alpha_i^{N\over 2}-\sum_{i=1}^4\beta_i^{N \over 2}\right).
\label{alphabeta}\end{eqnarray}
Here $\alpha_{i}$'s (resp. $\beta_{i}'s$) denote the roots of the 
algebraic equation
$R_{4}(X)=0$ (resp. $S_{4}(X)=0$). 

Note that the numerator of the right-hand-side of the addition formula 
(\ref{lem2}) is a polynomial $T_{16}(X)$ of order 16 which is
factored into a product of squares of $R_4(X)$ and $S_4(X)$. Thus each of 
the roots $\alpha_{i}$ and $\beta_{i}$ have a multiplicity 2 in $T_{16}(X)$.
On the other hand, if one
substitutes $z=1/8+(b+a\tau)/4, \hskip1mm
a,b \in {\bf Z}$ into (\ref{lem2}), the left-hand-side 
vanishes (recall the relation ${\cal P}(1/2+b+a\tau|\tau)=e_1(\tau)$).
Thus we recognize that $\{X=X(1/8+(b+a\tau)/4|\tau), \hskip1mm a,b=0,1,2,3\}$ 
are the solutions of the 
algebraic equation $T_{16}(X)=0$. From the definition
(\ref{Xdef}) $X(1/8+(b+a\tau)/4|\tau)=X(1/8+(4-b-1+(4-a)\tau)/4|\tau)$ and 
hence each of the roots has a multiplicity 2.
Then by reducing the range of $b$ to $b=0,1$, 
the roots $\{X=X(1/8+(b+a\tau)/4|\tau), \hskip1mm a=0,1,2,3,\hskip1mm b=0,1\}$ 
are in one to one
correspondence with $\{\alpha_i,\hskip1mm \beta_i, \hskip1mm i=1,2,3,4\}$.

Explicitly the zeros of $R_4(X)=0$ and $S_4(X)=0$ are given by
\ba
&&\left\{\begin{array}{ll}\alpha_j=\pi^2 
uv\big(uv-(u^2+v^2)\mp(u-v)\sqrt{u^2+v^2}
\big), & j=1,2, \\
\alpha_j=\pi^2 uv\big(uv+(u^2+v^2)\mp(u+v)\sqrt{u^2+v^2}\big), & j=3,4.
\end{array}\right.\label{sol1}\\
&&
\left\{\begin{array}{ll}\beta_j=\pi^2 uv\big(-uv-i(u^2-v^2)\pm(u-iv)
\sqrt{v^2-u^2}\big), &  j=1,2,\\
\beta_j=\pi^2 uv\big(-uv+i(u^2-v^2)\pm(u+iv)\sqrt{v^2-u^2}\big), 
& j=3,4,\end{array}\right.\label{sol2}
\ea

We find that
$\{\alpha_i, i=1,2,3,4\}=\{X(1/8+(b+a\tau)/4|\tau), (a=0,2,b=0,1)$\},
$\{\beta_i, i=1,2,3,4\}=\{X(1/8+(b+a\tau)/4|\tau), (a=1,3,b=0,1)\}$.

Thus we finally prove
\begin{theorem}
\begin{eqnarray}
&&\sigma({\cal L}M_{N}^{4})(\tau)=
-\frac{(-1)^{{N\over 2}}}{4}\sum_{a=0}^{3}
\sum_{b=0}^{3}(-1)^{a}\biggl(\frac{\theta_{2}
(\frac{1}{8}+\frac{b+a\tau}{4}|\tau)}
{\theta_{1}(\frac{1}{8}+\frac{b+a\tau}{4}|\tau)}\biggr)^{N}
\end{eqnarray}
\end{theorem}
This establishes the equivalence of LG theory with geometry for the 
case $k=4$.

By substituting (\ref{sol1}),(\ref{sol2}) into (\ref{alphabeta}) we obtain the 
results (\ref{LGk4}) 
of section 3.

\medskip

{\bf Remarks}\\

The $k=2$ case is easy and we leave this case to the reader.
We can prove the $k=6$ version of Theorem 1 along the same way, using 
the explicit calculation of the addition formula for $X(6z|\tau)$. The case 
of $\hat{A}$-genus can also be proved using the same method. In general 
the proof of 
the equivalence of LG theory with geometry is reduced to the proof of the 
analogue of Lemma 2 for general even values of $k$. 

In the case of odd $k$ we lose the factorization of the polynomial in
the numerator of the right-hand-side of the addition theorem. This necessarily
happens since the polynomial is degree $k^2$ which is not divisible by 2.
Then we can not express $\sigma({\cal L}M_N^k)$ as a residue integral in the 
$X$ variable.

\end{document}